\documentclass[12pt]{iopart}
\usepackage{bm}
\usepackage{graphicx}
\begin{document}
\title[Weak localization in quantum dots]{Weak localization in a system with a barrier: Dephasing and weak Coulomb blockade}
\author{Dmitri S. Golubev and Andrei D. Zaikin}
\address{
Forschungszentrum Karlsruhe, Institut f\"ur Nanotechnologie,
76021 Karlsruhe, Germany\\
I.E.Tamm Department of Theoretical Physics, P.N.Lebedev
Physics Institute, \\ 119991 Moscow, Russia
}

\begin{abstract}
We non-perturbatively analyze the effect of electron-electron interactions on
weak localization (WL) in relatively short metallic conductors with a tunnel
barrier. We demonstrate that the main effect of interactions is electron
dephasing which persists down to $T=0$ and yields suppression
of WL correction to conductance
below its non-interacting value. Our results may account
for recent observations of low temperature saturation of the electron
decoherence time in quantum dots.
\end{abstract}

\pacs{72.10.-d}
\pacs{71.30.+h}
\pacs{71.10.-w}
\submitto{\NJP}
\maketitle

Electrons propagating in a disordered conductor get scattered and
interfere. This quantum interference is possible only as long as
the electron wave functions remain coherent. In any realistic
situation, however, interactions between electrons and with other
degrees of freedom may limit phase coherence and, hence, reduce
electrons ability to interfere. The interplay between scattering,
quantum coherence and interactions yields a rich variety of
non-trivial effects and significantly impacts electron transport
in disordered conductors.

The so-called weak localization (WL) correction to the conductance
of a disordered system $G_{WL}$ is most sensitive to electron
coherence and is known to arise from interference of pairs of
time-reversed electron paths \cite{CS}. In a system of two
scatterers separated by a cavity (quantum dot) and in the absence
of interactions this correction can be directly evaluated
\cite{B}. The effect of electron-electron interactions can be described
in terms of fluctuating voltages. Provided the voltage drops only across the
barriers and not inside the cavity electron-electron interactions yield
energy dependent logarithmic renormalization of the dot channel transmissions
\cite{GZ01,BN} but {\it do not} cause any dephasing
\cite{GZ041,Brouwer}. The latter result can easily be understood
if one observes that the voltage-dependent random
phase acquired by the electron wave function $\Psi$ along any path
turns out to be the same as that for its time-reversed
counterpart. Hence, in the product $\Psi\Psi^*$ these random
phases cancel each other exactly and quantum coherence of
electrons remains preserved.

It is important, however, that this cancellation occurs only in
the case of two scatterers, whereas in a system of three or more
scatterers the situation is entirely different. Consider, e.g.,
a system of two quantum dots depicted in Fig. 1 and again assume
that fluctuating voltages are concentrated at the barriers. The
phase factor accumulated along the path (see Fig. 1) which crosses
the central barrier twice (at times $t_i$ and $t>t_i$) and returns
to the initial point (at a time $t_f$) is
$e^{i[\varphi^+(t_i)-\varphi^+(t)]}$, where $\dot\varphi^+/e=V(t)$ is
the fluctuating voltage across the central barrier. Similarly, the
phase factor picked up along the time-reversed path reads
$e^{i[\varphi^+(t_f+t_i-t)-\varphi^+(t_f)]}$. Hence, the overall
phase factor acquired by the product $\Psi\Psi^*$ for a pair of
time-reversed paths is $\exp (i \Phi_{\rm tot})$, where $\Phi_{\rm
tot}(t_i,t_f,t)=\varphi^+(t_i)-\varphi^+(t)-\varphi^+(t_f+t_i-t)+\varphi^+(t_f)$.
Averaging over phase fluctuations, which for simplicity are
assumed Gaussian, we obtain
\begin{eqnarray}
&&\left\langle e^{i \Phi_{tot}(t_i,t_f,t) } \right\rangle
=\,e^{-\frac{1}{2} \left\langle \Phi_{tot}^2(t_i,t_f,t)
\right\rangle} \nonumber\\ &&
=\,e^{-2F(t-t_i)-2F(t_f-t)+F(t_f-t_i)+F(t_f+t_i-2t)},
\label{phase}
\end{eqnarray}
where we defined the phase correlation function
\begin{equation}
F(t)=\langle (\varphi^+(t)-\varphi^+(0))^2\rangle /2. \label{F}
\end{equation}
Should this function grow with time the electron phase coherence
decays and $G_{WL}$ gets suppressed below its non-interacting
value.

\begin{figure}
\includegraphics[width=8.5cm]{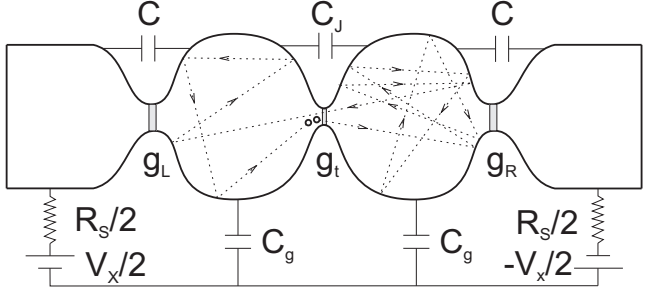}
\caption{Two quantum dots separated by a tunnel barrier and
connected to the battery $V_x$ via an Ohmic shunt resistor $R_S$.}
\end{figure}

The above arguments are not specific to the system of three
barriers and after proper generalization can be applied to
virtually any disordered conductor. At the same time, these
arguments are not yet sufficient to quantitatively describe the
decoherence effect of electron-electron interactions for two
important reasons: (i) fluctuating voltages are treated as
external (classical) fields rather than quantum fields produced
internally by fluctuating electrons and (ii) Fermi statistics is
not yet accounted for. Below we
will cure both these problems and non-perturbatively evaluate WL
correction $G_{WL}$ for a metallic system with a tunnel barrier
and (at least) two more scatterers in the presence of
electron-electron interactions which turn out to reduce phase
coherence of electrons at any temperature down to $T=0$.

We will consider a system with a tunnel barrier with dimensionless
conductance $g_t$, which separates two sufficiently short
disordered metallic conductors with Thouless energies $E_{\rm Th}$
and dimensionless conductances $g_{L,R}\gg 1, g_t$. This system is
described by the Hamiltonian
\begin{equation}
\hat {\bm H} =\hat {\bm H}_L+\hat {\bm H}_R+\hat {\bm T} +\hat {\bm H}_{em},
\label{HT}
\end{equation}
where $\hat {\bm H}_{L,R}=\sum_{\alpha=\uparrow,\downarrow}
\int_{L,R} d^3{\bm r}\,\hat\Psi_{\alpha;L,R}^\dagger({\bm r}) \hat
H_{L,R} \hat\Psi_{\alpha; L,R}({\bm r})$ is the Hamiltonian of the
left (right) lead, $\hat H_{L,R}=-\frac{\nabla^2}{2m_{L,R}}-\mu
+U_{imp}({\bm r})$ is the single electron Hamiltonian in the left
(right) lead, and $\hat T=\sum_{\alpha=\uparrow,\downarrow}
\int_{J}d^2{\bm r}\left [ t({\bm r})e^{-i\hat\varphi(t)}
\hat\Psi_{\alpha;L}^\dagger({\bm r})
 \hat\Psi_{\alpha; R}({\bm r})+c.c.\right]$ is the tunnel Hamiltonian.
Here $\hat\varphi$ is the phase operator, which is related to the
voltage drop across the junction $\hat\varphi(t)=\int^t_{t_0} dt'
e\hat V(t')$, and the ${\bm r}$-integration runs over the junction
area. Finally, $\hat{\bm H}_{em}\propto \hat V^2$ is the quadratic
Hamiltonian of electromagnetic fields, the precise form of which
depends on the circuit configuration and will not be specified
here.

Following the standard procedure we integrate out
fermionic degrees of freedom and arrive at the effective action
$iS=2\,{\rm Tr}\,\ln\left[\check G^{-1}\right]$, where $\check G$
is the Green-Keldysh function for our system. Expanding the action
in powers of the tunnel Hamiltonian we obtain
$S=S_{L,R}+S^{(1)}_t+S^{(2)}_t+...$, where the term $S_{L,R}$
describes the action of the left and right conductors, $S^{(1)}_t$
is Ambegaokar-Eckern-Sch\"on action \cite{SZ} and
\begin{eqnarray}
 iS^{(2)}_t&=& -\sum_{i,j,k,l=F,B}\int dt_1\dots dt_4 \int_J d\bm{x}_1\dots d\bm{x}_4\,
\nonumber\\ &&\times\, \check G_{L,ij}(X_1; X_2)(-1)^j
e^{-i\varphi_j(t_2)}t(\bm{x}_2) \times\, \check
G_{R,jk}(X_2;X_3)(-1)^k e^{i\varphi_k(t_3)}t(\bm{x}_3) \nonumber\\
&&\times\, \check G_{L,kl}(X_3; X_4)(-1)^l
e^{-i\varphi_l(t_4)}t(\bm{x}_4) \times\, \check
G_{R,li}(X_4;X_1)(-1)^i e^{i\varphi_i(t_1)}t(\bm{x}_1). \label{S2}
\end{eqnarray}
Here $X=(t,\bm{x})$, $\varphi_{F(B)}$ is the phase variable on the
forward (backward) branch of the Keldysh contour, $\check G_{r}$
are $2\times 2$ matrix Green-Keldysh functions in the left and
right conductors ($r=L,R$) and we use the convention
$(-1)^{F}=-1$, $(-1)^B=1$. We assume that $\check G_{L,R}$ have
the equilibrium form $\check G_r=G_r^R\check F_1-\check F_2G_r^A$,
where $G^{R,A}_r$ are retarded and advanced Green functions,
$$\check F_1 (E)= \left( \begin{array}{cc} h(E)& -f(E)\\ h(E) &
-f(E)\end{array}\right),
$$
$$
\check F_2 (E)= \left(
\begin{array}{cc} f(E)& f(E)\\ -h(E) &
-h(E)\end{array}\right),$$ $f(E)$ is the Fermi function and
$h(E)=1-f(E)$.

Our next step amounts to averaging the products of retarded and
advanced propagators in the action (\ref{S2}) over disorder in
each conductor separately. We have (see, e.g., \cite{Tan})
\begin{eqnarray}
\langle  G^R_L(X_1,X_2)G^A_L(X_3,X_4) \rangle =
\langle G^R_L(X_1,X_2)\rangle \langle G^A_L(X_3,X_4)\rangle
\nonumber\\
+\,2\pi N_L w(|\bm{r}_1-\bm{r}_4|)w(|\bm{r}_2-\bm{r}_3|) \times\,
{\cal
D}_{L}\left(t_1-t_2;\frac{\bm{r}_1+\bm{r}_4}{2},\frac{\bm{r}_2+\bm{r}_3}{2}\right)
\nonumber\\
\times \delta(t_1-t_2+t_3-t_4)  +\,2\pi N_L
w(|\bm{r}_1-\bm{r}_3|)w(|\bm{r}_2-\bm{r}_4|) \nonumber\\ \times\,
{\cal
C}_{L}\left(t_1-t_2;\frac{\bm{r}_1+\bm{r}_3}{2},\frac{\bm{r}_2+\bm{r}_4}{2}\right)
\delta(t_1-t_2+t_3-t_4) ,
\end{eqnarray}
where $N_L$, ${\cal D}_L(t,\bm{r},\bm{r}')$ and
${\cal C}_L(t,\bm{r},\bm{r}')$ are
respectively the density of states, the diffuson and the
Cooperon in the left conductor,
$w(r)=e^{-r/2l_e}\sin k_Fr/ k_Fr$, $k_F$ and $l_e$ are
respectively the Fermi wave vector and  elastic mean free path.
The same averaging procedure applies to the right conductor.

Finally we assume that the transmission amplitude $t(\bm{x})$ is
random, quickly oscillating real function. Averaging over these
oscillations yields
$\overline{t(\bm{x})t(\bm{y})}=\delta(\bm{x}-\bm{y}){g_t(\bm{x})}/{8\pi^2
N_LN_R}$, where $g_t(\bm{x})$ is the local conductance of the
barrier. After all these steps Eq. (\ref{S2}) reduces to a sum of
different terms. Here we will select only the terms responsible
for weak localization which involve the product of two Cooperons
${\cal C}_L$ and ${\cal C}_R$. Collecting all such contributions we
obtain
\begin{eqnarray}
&& iS_{WL}=-i\int dt_1\dots dt_4 \int d\tau_1 d\tau_2 \int_J d{\bm
x}d{\bm y} \nonumber\\ &&\times\, \frac{g_t({\bm x}) g_t({\bm
y})}{4\pi^2 N_LN_R}\;{\cal C}_{L}(t_1-\tau_1,{\bm y},{\bm x}){\cal
C}_{R}(t_2-\tau_2,{\bm x},{\bm y}) \times\, e^{i\Phi
(t_1,...,t_4)}\sin\frac{\varphi^-(t_1)}{2} \nonumber\\ && \times\,
\left[  h(\tau_1-t_2) e^{-i\frac{\varphi^-(t_2)}{2}}
 +  f(\tau_1-t_2) e^{i\frac{\varphi^-(t_2)}{2}} \right]
\nonumber\\ && \times\, \left[
h(\tau_2-t_3)e^{i\frac{\varphi^-(t_3)}{2}}f(t_1+t_3-t_4-\tau_1)
\right. \nonumber\\ && \left. -\, f(\tau_2-t_3)
e^{-i\frac{\varphi^-(t_3)}{2}}h(t_1+t_3-t_4-\tau_1) \right]
\nonumber\\ &&\times\, \left[
e^{-i\frac{\varphi^-(t_4)}{2}}f(-t_1+t_2+t_4-\tau_2) +\,
e^{i\frac{\varphi^-(t_4)}{2}}h(-t_1+t_2+t_4-\tau_2)\right]
\nonumber\\ && +\,\big\{ L\leftrightarrow R, \varphi^\pm\to
-\varphi^\pm\big\}. \label{SWL}
\end{eqnarray}
Here we defined ``classical'' $\varphi^+=(\varphi_F+\varphi_B)/2$ and
``quantum'' $\varphi^-=\varphi_F-\varphi_B$ phases and introduced
$$
\Phi (t_1,...,t_4)=\varphi^+(t_1)-\varphi^+(t_2)+\varphi^+(t_3)-\varphi^+(t_4)
$$
and $f(t)=\int (dE/2\pi)\,f(E)e^{-iEt} \equiv\delta (t)-h(t)$.
The action (\ref{SWL}) fully accounts for the effects
of electron-electron interactions on WL via the fluctuating phases
$\varphi^{\pm}$.

In order to find the WL correction to the current across the
central barrier $I_{WL}$ we make use of the following general formula
\begin{equation}
I_{WL}=ie\int D^2\varphi^{\pm}\, \frac{\delta iS_{WL}[\varphi^{\pm}]}{\delta \varphi^-}\,
 e^{iS_{L,R}+iS^{(1)}_{\rm t}}.
\label{I}
\end{equation}
In the limit $g_{L,R} \gg 1, g_t$ this integral remains Gaussian
in $\varphi^{\pm}$ at all relevant energies and can easily be
performed. The effective expansion parameter in this case is
$g_t^2/g_Lg_R\ll 1$. Combining Eqs. (\ref{SWL}) and (\ref{I})
and introducing the average voltage at the barrier
$V$ we find
\begin{eqnarray}
 && I_{WL} =\frac{e}{8\pi^3 N_LN_R}\,{\rm Re}\,\int_J d{\bm x}d{\bm y}
\; g_t({\bm x}) g_t({\bm y})\; \int dEd\omega_1 d\omega_2
d\omega_3\, \nonumber\\ &&\times\, 
C_{R}(-\omega_2,{\bm x},{\bm y})C_{L}(-\omega_3,{\bm y},{\bm x}) \times\,
h(E-\omega_2)f(E+eV+\omega_3-\omega_1) \nonumber\\ &&\times\, \big[
f(E+eV-\omega_1)h(E)P_1(\omega_1,\omega_2,\omega_3)
+\,f(E+eV-\omega_1)f(E)P_2(\omega_1,\omega_2,\omega_3) \nonumber
\\ && +\,h(E+eV-\omega_1)h(E)P_2(\omega_1,\omega_3,\omega_2)
\nonumber\\ &&
+\,h(E+eV-\omega_1)f(E)P_3(\omega_1,\omega_2,\omega_3) \big]
-\,\big\{ V\to -V \big\}. \label{IWL1}
\end{eqnarray}
Here $C_{L,R}$  and $P_{j}$ ($j=1,2,3$) are the Fourier transforms of
respectively the Cooperons ${\cal C}_{L,R}(t)$ and the functions
\begin{equation}
{\cal P}_{j}(t_1,t_2,t_3)=\exp [-{\cal F}(t_1,t_2,t_3)]{\cal
Q}_{j}(t_1,t_2,t_3), \label{calP}
\end{equation} where
\begin{eqnarray}
&& {\cal
F}=F(t_1+t_3)+F(t_3)+F(t_1+t_2)+F(t_2)\nonumber\\ &&-F(t_1+t_2+t_3)-F(t_2-t_3)
\label{calF}
\end{eqnarray}
and $F(t)=\langle (\hat\varphi(t)-\hat\varphi (0))^2\rangle /2$
coincides with the phase correlation function (\ref{F}). The terms
${\cal Q}_j$ read
\begin{eqnarray}
&& {\cal Q}_1= e^{-i\left[K(t_2)+K(t_3)+K(|t_2-t_3|)\right]}
\nonumber\\ &&\times\,
\big\{2e^{i\left[K(|t_1+t_2+t_3|)+K(t_1+t_3)+K(t_1+t_2)\right]}
-\,e^{i\left[K(t_1+t_2+t_3)+K(|t_1+t_3|)+K(|t_1+t_2|)\right]}
\big\},
\nonumber\\
&& {\cal Q}_2=e^{i\left[K(|t_1+t_2+t_3|)-K(t_2)-K(|t_3|)\right]}
e^{i\left[K(t_1+t_3)-K(|t_1+t_2|)-K(t_3-t_2)\right]},
\nonumber\\
&& {\cal Q}_3= e^{i\left[K(t_1+t_2+t_3)-K(|t_2|)-K(|t_3|)\right]}
e^{-i\left[K(t_1+t_3)+K(t_1+t_2)-K(|t_3-t_2|)\right]},
 \label{Pj}
\end{eqnarray}
where $K(t)=i\left\langle \big[\hat\varphi(0),\hat\varphi(t)\big]
\right\rangle$ is the response function. Eqs.
(\ref{IWL1})-(\ref{Pj}) represent the central result of our paper.
They fully determine WL correction to the current in our system.
The non-interacting result is reproduced by the first two lines of
Eq. (\ref{IWL1}) before the square brackets, while the terms in
the square brackets exactly account for the effect of
interactions. The same result follows from the non-linear
$\sigma$-model approach \cite{GZ06}.

Our result demonstrates that the whole effect of electron-electron
interactions is encoded in two different correlators of
fluctuating phases $F(t)$ and $K(t)$. These correlation functions
are well familiar from the so-called $P(E)$-theory \cite{SZ,IN}.
They read
\begin{eqnarray}
F(t)&=&e^2\int\frac{d\omega}{2\pi}\,\omega\coth\frac{\omega}{2T}\,
{\rm Re}\big[Z(\omega)\big]\frac{1-\cos\omega t}{\omega^2},
\label{FFF}\\
K(t)&=&e^2\int\frac{d\omega}{2\pi}\,{\rm Re}\big[Z(\omega)\big]
\frac{\sin\omega t}{\omega}, \label{KKK}
\end{eqnarray}
where $Z(\omega)$ is an effective impedance ``seen'' by the
central barrier. Both functions (\ref{FFF}) and (\ref{KKK}) are
purely real and, hence, $|{\cal Q}_j|\leq 1$. At times $
\tau_{RC}<|t|< 1/E_{\rm Th}$ (an effective $RC$-time $\tau_{RC}$
will be defined later) we obtain $F(t)\simeq \frac{2}{g_Z}
\left(\ln\left|\frac{\sinh\pi Tt}{\pi
T\tau_{RC}}\right|+\gamma\right)$ and $K(t)\simeq
\frac{\pi}{g_Z}\,{\rm sign}\, t$, where $g_Z = 2\pi /e^2
Z(0)=g_0+g_t$, $g_0^{-1}=g_L^{-1}+g_R^{-1}+e^2R_S/2\pi$ and $\gamma \simeq
0.577$ is the Euler constant. We
observe that while $F(t)$ grows with time at any temperature
including $T=0$, the function $K(t)$ always remains small in the
limit $g_Z\gg 1$ considered here. Hence, the combination
(\ref{calF}) should be fully kept in the exponent of (\ref{calP})
while the correlator $K(t)$ can be safely ignored
in the leading order in $1/g_Z$. Then all ${\cal Q}_j \equiv 1$, the
Fermi function $f(E)$ drops out from the result and we get
$I_{WL}= G_{WL}V$, where
\begin{eqnarray}
 G_{WL}&=& -\frac{e^2}{8\pi^3N_LN_R}\int dt_2dt_3 \int_J
d^2\bm{x}d^2\bm{y}\, \nonumber\\ &&\times\, g_t(\bm{x})g_t(\bm{y})
{\cal C}_L(t_2,\bm{y},\bm{x}) {\cal C}_R(t_3,\bm{x},\bm{y})
\nonumber\\ &&\times\, e^{-2F(t_2)-2F(t_3)+F(t_2+t_3)+F(t_2-t_3)}. \label{GWLF}
\end{eqnarray}
Identifying $t_2=t_f-t$ and $t_3=t-t_i$ we observe that the
exponent in the third line of Eq. (\ref{GWLF}) exactly coincides
with the expression (\ref{phase}) derived from simple
considerations involving electrons propagating along time-reversed
paths in an external fluctuating field. Thus, in the leading order
in $1/g_Z$ the WL correction $G_{WL}$ is affected by
electron-electron interactions via dephasing produced {\it only}
by the ``classical'' component $\varphi^+$ of the fluctuating
field which mediates such interactions. Fluctuations of the
``quantum'' field $\varphi^-$ turn out to be irrelevant for
dephasing and may only cause a (weak) Coulomb blockade correction
to be considered below.

It is worthwhile to point out that a similar conclusion was
previously reached for spatially
extended disordered conductors within a different approach
\cite{GZ}. We also note that a close relation between the results \cite{GZ} and
the $P(E)$-theory \cite{SZ,IN} was already
demonstrated earlier \cite{GZ99}. Our present
results make this relation even more transparent.

Our further calculation is concentrated on a system of two
(identical) dots depicted in Fig. 1. For simplicity the outer barriers
are supposed to be open, $g_{L,R}=g \gg 1, g_t$ and $R_S \to 0$. Then
the Cooperons  take a simple form
${\cal C}_{L,R}(t)={e^{-t/\tau_D}}/{\cal V}$, where ${\cal V}$ and
$\tau_D$ are respectively the dot volume and dwell time. We also define
the effective impedance seen by the central tunnel junction
\begin{eqnarray}
Z(\omega)= i\frac{4\pi}{e^2g}\frac{1}{\tau_D+\tau_{RC}}
\left(\frac{\tau_D}{\tau_{RC}}\frac{1}{\omega+\frac{i}{\tau}}
+\frac{1}{\omega+i0}\right),
\end{eqnarray}
with the real part
\begin{eqnarray}
{\rm Re}\, Z(\omega)&=&\frac{4\pi}{e^2 g}\bigg[\frac{\tau^2}{\tau_{RC}^2}\frac{1}{1+\omega^2\tau^2}
+\frac{\pi}{\tau_D+\tau_{RC}}\delta(\omega)\bigg],
\label{imped}
\end{eqnarray}
where $1/\tau =1/\tau_D+1/\tau_{RC}$, $\tau_{RC}=\pi /gE_C$,
$E_C=e^2/2(C+C_g+2C_J)$ and $C$, $C_J$ and $C_g$ are the
capacitances of respectively left (right) barriers, the central
junction and the gate electrode. Substituting the Cooperons ${\cal
C}_{L,R}(t)$ and the correlator $F(t)$ (\ref{FFF}), (\ref{imped})
into Eq. (\ref{GWLF}) we observe that contribution of
$\delta(\omega)$ in Eq. (\ref{imped}) drops out. Performing the
time integrals we arrive at the final expression for the WL
correction $G_{WL} (T)$ in the presence of electron-electron
interactions:
\begin{eqnarray}
 G_{WL}&=& -\frac{e^2 g_t^2\delta^2}{8\pi^3}\int dt_2dt_3\,
e^{-(t_2+t_3)/\tau_D} e^{-2F(t_2)-2F(t_3)+F(t_2+t_3)+F(t_2-t_3)},
\label{GWLF1}
\end{eqnarray}
where $\delta$ is the dot mean level spacing.
This result is plotted in Fig. 2a demonstrating that
interactions suppress $G_{WL} (T)$ below
its non-interacting value \cite{GZ06} $G^{(0)}_{WL}=-2e^2
g_t^2/\pi g^2$.

\begin{figure}
\begin{tabular}{cc}
(a) & (b) \\
\includegraphics[width=6.15cm]{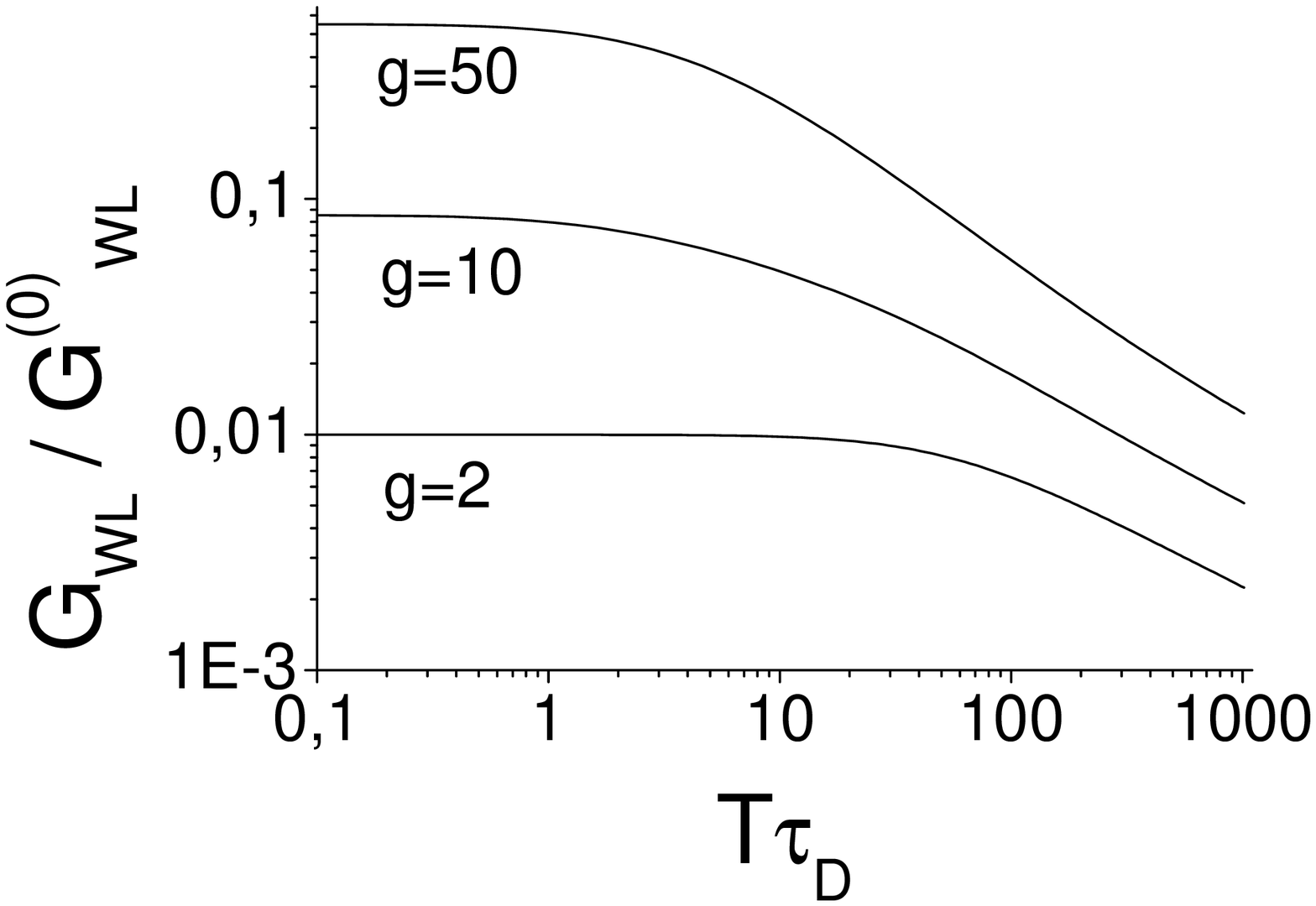} &
\includegraphics[width=5.25cm]{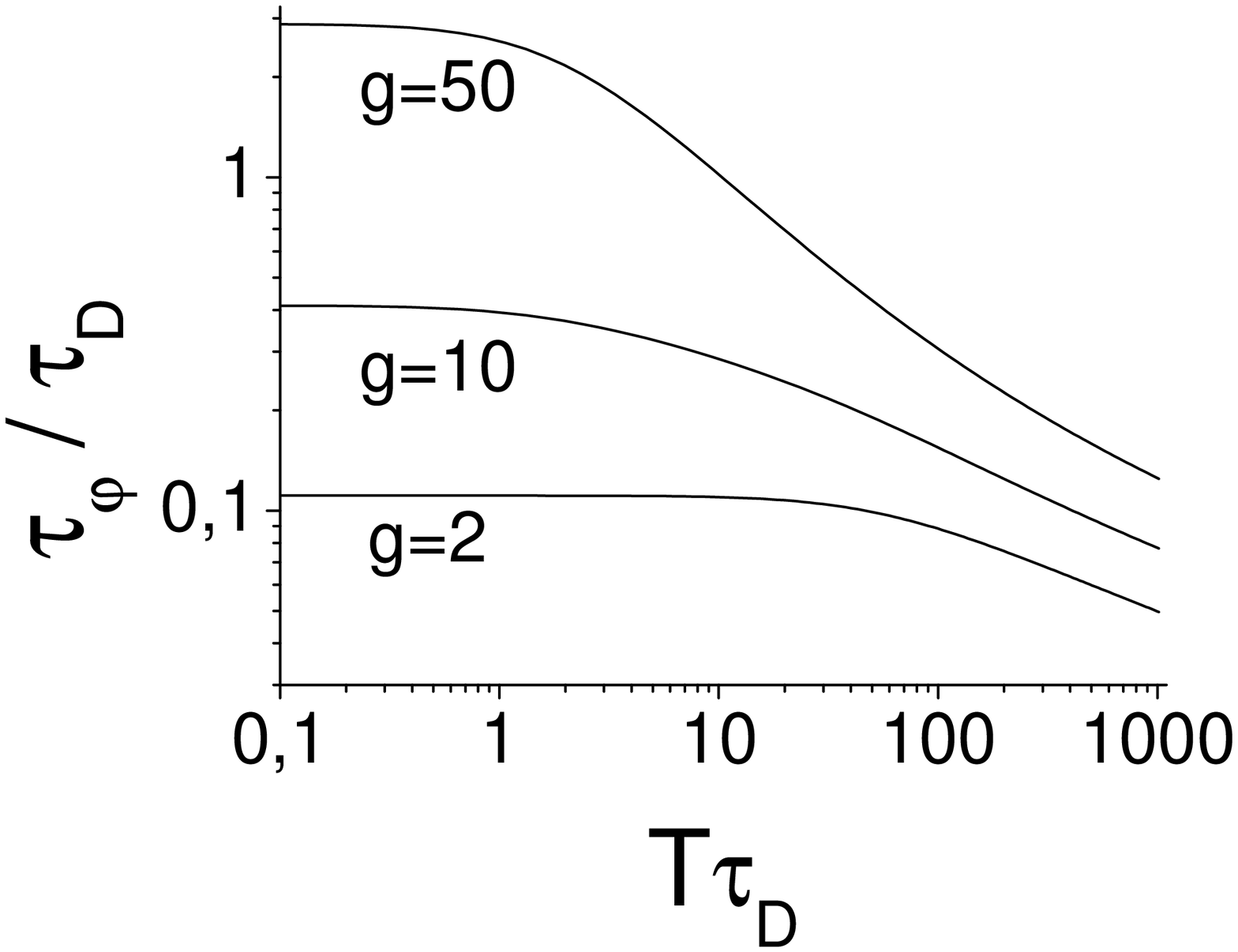}
\end{tabular}
\caption{Temperature dependence of WL correction $G_{WL}$ (a) and
dephasing time $\tau_{\varphi}$ (b) for $\tau_D/\tau_{RC}=100$. }
\end{figure}

Let us define $u=\tau_D/\tau_{RC}=4E_C/\delta$  and consider the limit of
metallic dots $u\gg 1$. At $T\tau_D \leq 1$ the WL correction saturates to
\begin{eqnarray}
{G_{WL}}/{|G_{WL}^{(0)}|}&\simeq & -\left(2/u\right)^{{8}/{g}},\;\; g\geq 8,
\nonumber\\
{G_{WL}}/{|G_{WL}^{(0)}|}&\simeq & -g/2u,\;\; 1\leq g\leq
8,
\label{t0}
\end{eqnarray}
whereas at $g/\tau_D\leq T\leq
1/\tau_{RC}$ and for $g\geq 8$ we find
\begin{eqnarray}
\frac{G_{WL}}{|G_{WL}^{(0)}|}&\simeq & -\left(\frac{g}{4}-2\gamma\right)
\frac{\left({2\pi}/{u}\right)^{{8}/{g}}}{(\pi T\tau_D)^{1-8/g}}.
\end{eqnarray}
Let us phenomenologically define the electron decoherence time
$\tau_\varphi$ by taking the Cooperons in the form
${\cal C}_{L,R}(t)={e^{-t/\tau_D-t/\tau_\varphi}}/{\cal V}$ which yields
\cite{GZ06}
${G_{WL}}/{G_{WL}^{(0)}}=(1+\tau_D/\tau_\varphi)^{-2}$. Resolving
this equation for $\tau_\varphi$ we obtain
\begin{equation}
\tau_\varphi/\tau_D=\left(\sqrt{G_{WL}^{(0)}/G_{WL}}-1\right)^{-1},
\label{tphi}
\end{equation}
which yields $\tau_{\varphi}=g\tau_D/4\ln (2E_C/\delta )$
for $T\tau_D \leq 1$ and $g \gg 8\ln(u/2)$.
Eqs. (\ref{t0})-(\ref{tphi}), although not directly applicable
to a single quantum dot,
account for key features of the
dependence $\tau_{\varphi}(T)$ (Fig. 2b) observed in
various UCF experiments
\cite{Pivin,Huibers,Hackens} with quantum dots \cite{FN0}. At higher
temperatures we find $\tau_\varphi \propto T^{-\nu}$ with
non-universal $g$-dependent power $\nu \leq 1/2$, while at
lower $T \leq 1/\tau_D$ the electron decoherence time
$\tau_{\varphi}$ saturates to a constant in agreement with
the observations \cite{Pivin,Huibers,Hackens}. It was
pointed out \cite{Hackens} that the available experimental values
of $\tau_{\varphi}(0)$ scale as $\tau_{\varphi}(0)\approx \tau_D$
for a variety of dot sizes and dwell times $\tau_D$ varying by
$\sim 3$ decades. Our result (\ref{t0},\ref{tphi}) should be
consistent with this scaling provided at low $T$ the right-hand
side of Eq. (\ref{tphi}) remains of order one.

Note that the phenomenological
definition of $\tau_\varphi$ is identical to that used before in
Ref. \cite{GZ06} where we also demonstrated that for an arbitrary array of
quantum dots our expression for the weak localization correction
determines the system magnetoconductance if we substitute $1/\tau_\varphi
\to 1/\tau_\varphi +1/\tau_H$, where $\tau_H \propto 1/H^2$ is the electron
dephasing time due to the external magnetic field $H$. Thus, our definition of
$\tau_\varphi$ is fully consistent with the standard procedure of extracting
the electron dephasing time from the magnetoconductance curves. Furthermore,
it is straightforward to demonstrate \cite{GZ06} that, e.g., in the case of
quasi-1d arrays of quantum dots our definition for $\tau_\varphi$ just yields
the standard result for the magnetoconductance of a diffusive wire,
cf. Eq. (60) of Ref. \cite{GZ06}.

We also would like to emphasize that there
exists no contradiction between the definition of $\tau_\varphi$ adopted here
and the fact that no dephasing occurs for electon paths confined within
a single quantum dot, as discussed in the beginning of our manuscript. As
it was demonstrated, electron dephasing occurs as soon as time-reversed paths
cross the central barrier twice and return to the initial point inside the dot
(see Fig. 1). In the presence of fluctuating electromagnetic potentials
(dropping across the central barrier) the forward path and its time-reversed
counterpart pick up different random phases. After averaging over both
fluctuating fields and electron paths one arrives at a decaying in time
contribution to the Cooperons which is just captured
by our phenomenological definition. Of course, other definitions
of $\tau_\varphi$ can also be
employed. However, our basic conclusion about non-vanishing electron dephasing
by electron-electron interactions down to $T \to 0$ will not be sensitive
to any particular definition of $\tau_\varphi$, since this conclusion is based
on the result (\ref{GWLF}) demonstrating the interaction-induced
suppression of the WL correction to
conductance (as well as of the magnetoconductance, cf. Ref. \cite{GZ06})
at any temperature including $T=0$. The basic physics behind this result
is exactly the same as that already elucidated by the well known
$P(E)$-theory \cite{SZ,IN}: tunneling electrons can exchange energies with
an effective electromagnetic environment. This process results in broadening
of the distribution function for such electrons even at $T=0$ which inevitably
yields electron dephasing.

Finally, it is instructive to establish the relation to the
ordinary perturbation theory in the interaction which is
reproduced by formally expanding our exact result (\ref{IWL1}) to
the first order in $Z(\omega)$. We obtain
\begin{equation}
I_{WL}=G_{WL}^{(0)}V+\delta I_{WL}^{F}(V)+\delta I_{WL}^K(V),
\label{pert}
\end{equation}
where
\begin{eqnarray}
&& \delta I_{WL}^F=-\frac{e^3g_t^2\delta^2}{8\pi^3} eV
\int\frac{d\omega}{2\pi}\,\frac{{\rm Re}\,Z(\omega)}{\omega}
\coth\frac{\omega}{2T}
\nonumber\\ &&\times\,
\big[2 C_L(0) C_R(\omega)+2 C_L(\omega)C_R(0)
-2 C_L(0)C_R(0)
\nonumber\\ &&
-\, C_L(\omega) C_R(\omega)- C_L(-\omega) C_R(\omega)
\big],
\label{WLF}
\end{eqnarray}
\begin{eqnarray}
&& \delta I_{WL}^K=\frac{e^3 g_t^2\delta^2}{16\pi^3}
\int\frac{d\omega}{2\pi}\frac{W(\omega,V)}{\omega}\big\{
\,{\rm Re}\,Z(\omega)
\nonumber\\ &&\times\,
\big[ 2C_L(0) C_R(\omega)
+2 C_L(\omega) C_R(0) - C_L(-\omega) C_R(\omega)\big]
\nonumber\\ &&
+\, i\,{\rm Im}\, Z(\omega)\; C_L(\omega)C_R(\omega)
\big\}.
\label{WLK}
\end{eqnarray}
Here we defined the function
$$
W=(\omega+eV)\coth\frac{\omega+eV}{2T}
-(\omega-eV)\coth\frac{\omega-eV}{2T},
$$
and Fourier transformed Cooperons
$C_L(\omega)=C_R(\omega)=\tau_D/(1-i\omega\tau_D)$. The two terms
$\delta I_{WL}^F$ and $\delta I_{WL}^K$ are linear in respectively
$F(t)$ and $K(t)$.

Exactly the same results
(\ref{pert})-(\ref{WLK}) are reproduced from  the first order
diagrammatic perturbation theory in the interaction. In order to
observe the equivalence of the two approaches one should keep in
mind that $F(t)$ is proportional to the Keldysh component of the
photon Green function, while $K(t)$ is proportional to the
retarded photon Green function. One should also remember that the
photon Green function in our model is coordinate independent in
both quantum dots. One can actually demonstrate that the terms
$\propto G_L(\omega)G_R(0),$ $G_L(0)G_R(\omega)$ come from the
so-called "self-energy" diagrams, while the terms $\propto
G_L(-\omega)G_R(\omega)$ emerge from the "vertex" diagrams.

The term $\delta I_{WL}^K$ represents the Coulomb blockade
correction to $I_{WL}^{(0)}$ and is entirely different from the
dephasing term $\delta I_{WL}^F$. In contrast to the latter, the
term $\delta I_{WL}^K$ is non-linear in $V$ describing the
standard Coulomb offset at large $V$
and turning into
\begin{equation}
\delta I_{WL}^K/|I_{WL}^{(0)}| \sim 1/g\tau_DT
\label{wb}
\end{equation}
for $T\tau_D\geq 1$ in the linear in $V$ regime. Thus, the Coulomb
blockade correction remains small \cite{FN} in the metallic limit
$g\gg 1$. We also note that $\delta I_{WL}^K$ involves the
combination $1-2f(E)=\tanh (E/2T)$ which enters {\it only} in the
first order in the interaction. As in the case of spatially
extended conductors, at $T=0$ some terms contained in $\delta
I_{WL}^K$ partially cancel similar contributions to $\delta
I_{WL}^F$. This cancellation, however, remains incomplete and, as
demonstrated by our exact result, by no means implies the absence
of electron dephasing at $T\to 0$. More information on the debates
on low temperature decoherence by electron-electron interactions
can be obtained, e.g., from Refs. \cite{many} and further
references therein. Without going into details, we would only like
to emphasize that our present manuscript does not make any use of the
techniques introduced in our previous works on decoherence in disordered
conductors and, hence, is formally independent on those.

In summary, we have non-perturbatively treated the effect of
electron-electron interactions on weak localization  in relatively
short metallic conductors. The most significant effect of
interactions is electron decoherence which persists down to $T=0$
and -- in agreement with experiments \cite{Pivin,Huibers,Hackens}
-- yields saturation of $\tau_{\varphi}$ at $T \leq 1/\tau_D$.
The physics behind this effect is exactly the same as that discussed,
e.g., within the well known $P(E)$-theory \cite{SZ,IN}.
It is also worth pointing out that very recently
\cite{GZ07} we generalized our present approach to arbitrary
arrays of quantum dots and derived the expression for
$\tau_{\varphi 0}$ which describes both weakly and strongly
disordered conductors and quantitatively explains numerous
experimental data available to date. In the case of weakly
disordered conductors our results \cite{GZ07} match with those
derived previously \cite{GZ} by means of a different technique.

This work was supported in part by the EU Framework Programme
NMP4-CT-2003-505457 ULTRA-1D "Experimental and theoretical
investigation of electron transport in ultra-narrow 1-dimensional
nanostructures" and by RFBR Grant 06-02-17459.


\begin{thebibliography}{30}
\bibitem{FN0} Our results derived for a double dot system can also
  -- at least qualitatively -- apply to a single
dot provided fluctuating voltage is not strictly uniform inside the cavity.
\bibitem{FN} Somewhat more accurate treatment of the Coulomb blockade
  correction amounts to expanding the
exact result only in $K(t)$. Then one again arrives at Eq.
(\ref{wb}) with $I_{WL}^{(0)} \to I_{WL}$. One can also
demonstrate that at $T<1/\tau_D$ the Coulomb blockade correction
saturates. For $g \gg 1$ this correction remains unimportant down
to exponentially small energies below which weak Coulomb blockade
turns into strong at any non-zero $R_S$.
\end{thebibliography}
\end{document}